\documentclass[prd,aps,showkeys,prd,showpacs,superscriptaddress]{revtex4}

\usepackage{epsfig}
\usepackage{graphicx}
\usepackage{indentfirst}
\usepackage{amsmath}
\usepackage{amsfonts}
\usepackage{amssymb}
\usepackage{array}
\def\be{\begin{equation}}

\def\ee{\end{equation}}

\usepackage{breakurl}

\usepackage{epsfig}

\begin{document}

\title{Thin-Shell Wormholes in Neo-Newtonian Theory}

\author{Ali \"{O}vg\"{u}n}
\email{aovgun@gmail.com}

\affiliation{Instituto de F\'{\i}sica, Pontificia Universidad Cat\'olica de
Valpara\'{\i}so, Casilla 4950, Valpara\'{\i}so, Chile}

\affiliation{Physics Department, Eastern Mediterranean University, Famagusta,
Northern Cyprus, Turkey}

\author{Ines G. Salako}
\email{inessalako@gmail.com}

\affiliation{Institut de Mathematiques et de Sciences Physiques, Universit\'e de Porto-Novo, 01 BP 613, Porto-Novo, Benin}

\affiliation{D\'epartement de Physique, Universit\'e Nationale d’Agriculture, 01 BP 55, Porto-Novo, Benin}

\date{\today}

\begin{abstract}
In this paper, we constructed an acoustic thin-shell wormhole (ATW) under neo-Newtonian theory using the Darmois-Israel junction conditions. To determine the stability of the ATW by applying the cut-and-paste method, we found
the surface density and surface pressure of the ATW under neo-Newtonian hydrodynamics just after obtaining an analog acoustic neo-Newtonian solution.
We focused on the effects of the neo-Newtonian parameters by performing 
stability analyses using different types of fluids, such as a linear barotropic fluid (LBF), a Chaplygin fluid (CF), 
a logarithmic fluid (LogF), and a polytropic fluid (PF).
We showed that a fluid with negative energy is required at the throat to keep the wormhole stable. The ATW can be stable if
suitable values of the neo-Newtonian parameters $\varsigma$, $A$, and $B$ are chosen.
 
\end{abstract}

\maketitle
\section{Introduction}\label{section1}

Einstein's general theory of relativity is one of the towering achievements of twentieth-century theoretical physics and has contributed many important ideas to this field, such as the existence of black holes and
compact objects. The theory of general relativity has also revealed the existence of objects 
called wormholes that connect two different regions of the universe \cite{fl,er}. The pioneering work on wormholes was first performed by Morris and Thorne \cite{mth1,mth2}, and then Visser had the brilliant idea of building thin-shell wormholes \cite{mth2,mv1,mv2,mv3,is} to minimize the negative matter in the throat. Since Visser`s novel work, various thin-shell wormholes have been studied \cite{eiroa1,ayan1,lobo1,eiroa2,ali1,alii,rah,mazh1,ali,kimet,kimett,po,eiroa3,ali2,ern1,ern2,lemos1,lemos2,ayan2,ayan3,
myrzakulov,kim,peter1,sharifcy1,sharifcy2,sh1,sh2,sh3,baji,Xian,bilic,fagnic,visser}).

In this study, our aim is to construct acoustic thin-shell wormholes (ATWs) under neo-Newtonian theory. For this purpose, we briefly study analog gravity, which is a classical Newtonian treatment. The pressure is the main ingredient of general relativity; indeed, the Newtonian approach is valid only for pressure-less fluids, so that there is a neo-Newtonian generalization that can incorporate pressure effects. Neo-Newtonian theory gives a first-order correction, as a result, that is approximately similar to the exact general relativity \cite{velten,oli,fab}. In the literature, there are many applications of neo-Newtonian theory that provide interesting ways to study the effects of analog gravity, 
such as the Aharonov-Bohm (AB) effect caused by the acoustic geometry of a vortex in the fluid \cite{Fabris2013}.
McCrea in~\cite{McCrea} deduced the neo-Newtonian equations that were later refined in~\cite{Harrison}; \cite{Lima1997} later obtained a final expression for the equation of fluid that considers 
a perturbative treatment of the neo-Newtonian equations (see also \cite{velten,oli,fab}). Moreover, \cite{velten,oli,fab} studied acoustic black holes in the framework of neo-Newtonian hydrodynamics, and \cite{Salako:2015tja} analyzed the effect of neo-Newtonian hydrodynamics on the super-resonance phenomenon.

Analog gravity has led to a number of ideas such as an analog Schwarzschild metric solution known as a canonical
acoustic metric, the Painleve\textendash Gullstrand acoustic metric \cite{visser1}, a rotating analog
metric \cite{visser1,unruh2,visser2}, and analog Anti-de-Sitter (AdS) and de Sitter (dS) black hole solutions \cite{lib}.
In addition, following previous results, Nandi et al. \cite{nandi} introduced the concept of acoustic traversable wormholes, 
using the analogy of acoustic black holes. This technique was also used to investigate the nature of curvature singularities to study the light ray trajectories
in an optical medium, which are equivalent to the sound trajectories in the acoustic analog.
On the other hand, we note a series of studies that were carried out to calculate the quasi-normal modes,
the super-radiance, and the area spectrum \cite{saavedra1,saavedra2}.
 
  In this article, we study an ATW under neo-Newtonian hydrodynamics, which is a modification of the usual Newtonian theory
that correctly incorporates the effects of pressure. We concentrate on investigating the stability of the ATW using different types of gases, such as a linear barotropic fluid (LBF) \cite{kuf2,varela}, a Chaplygin fluid (CF) \cite{cg1,cg2,GCCG,GCCG1},
a logarithmic fluid (LogF) \cite{ali1,ali2}, and a polytropic equation of state for the fluid (PF) \cite{sarkar}. The paper is organized as follows:  In Sec. \ref{section2},
we review acoustic black holes under neo-Newtonian theory. In Sec. \ref{section4}, we construct 
the ATW and show that an exotic fluid with negative energy is required at the throat to keep the wormhole stable. In Sec. \ref{section5}, we investigate the stability analyses using the 
CF, LogF, and PF. In Sec. \ref{section6}, we discuss our results. 

\section{Acoustic Black Holes in neo-Newtonian theory} \label{section2}

 McCrea \cite{McCrea} and Harrison \cite{Harrison} developed
the basic foundations of neo-Newtonian theory in which the effects of pressure are considered contrary to the Newtonian theory.
In this section, we present a brief overview of neo-Newtonian hydrodynamics and introduce 
the acoustic black hole metric obtained in \cite{Fabris2013}.
First, we present the neo-Newtonian equations are given by~\cite{Fabris2013,McCrea,Harrison,rrrr,velten,oli,fab,carlos}
\begin{equation}
\partial_{t} \rho_i + \nabla\cdot(\rho_i\vec v) + p \nabla\cdot\vec{v} = 0\; 
\label{salako1bis}
\end{equation}
and
\begin{equation}\label{salako2}
\dot{ \vec{v}} + (\vec{v} \cdot  \nabla )\vec{v} = - \frac{ \nabla  p}{\rho + p}\,.
\end{equation}
Note that $\rho_i$ is an initial fluid density, $p$ is a pressure and $\vec v$ is a  {flow/fluid velocity}. 
The Eq.(\ref{salako1bis}) and Eq.(\ref{salako2}) are the continuity equation and the Euler 
equation  modified due to gravitational interaction, respectively. It is to be noted that the newtonian equations are recovered for a small pressure ($p\sim 0$).

 We assume that the fluid is barotropic, i.e.
$ p=p(\rho)$, inviscid and irrotational,  being the equation of state $p = k \rho^{n}$, with $k$ and $n$ constants. 
We write the fluid velocity as $ \vec{v}=-\nabla\psi $ where $ \psi $ is the velocity potential.
Now, we linearise the equations (\ref{salako1bis}) and (\ref{salako2}) by perturbating 
$ \rho $, $ \vec{v} $ and $ \psi $ as follows:

\begin{eqnarray}
\rho & = & \rho_{0}+\varepsilon\rho_{1}+0(\varepsilon^{2})\;,\\
\rho^{n} & = & \left[\rho_{0}+\varepsilon\rho_{1}+0(\varepsilon^{2})\right]^{n}\approx\rho_{0}^{n}+n\varepsilon\rho_{0}^{n-1}\rho_{1}+...\;,\\
\vec{v} & = & \vec{v}_{0}+\varepsilon\vec{v}_{1}+0(\varepsilon^{2}),\\
\psi & = & \psi_{0}+\varepsilon\psi+0(\varepsilon^{2})\;,\label{salako80'}
\end{eqnarray}
{where $\rho$ is the fluid density. Then the wave equation becomes 
\begin{eqnarray}
 & - & \partial_{t}\Big\{ c_{s}^{-2}\rho_{0}\Big[\partial_{t}\psi+\Big(\frac{1}{2}+\frac{\varsigma}{2}\Big)\vec{v}_{0}.\nabla\psi\Big]\Big\}+\nabla\cdot\Big\{-c_{s}^{-2}\rho_{0}\vec{v_{0}}\Big[\Big(\frac{1}{2}+\frac{\varsigma}{2}\Big)\partial_{t}\psi +\varsigma\vec{v_{0}}.\nabla\psi\Big]+\rho_{0}\nabla\psi\Big\}=0,\label{salako12bis}
\end{eqnarray} where $\varsigma=1+kn\rho_{0}^{n-1}$,
that can be given as 
\begin{eqnarray}
\partial_{\mu}(f^{\mu\nu}\partial_{\nu}\psi)=0.\label{eqf}
\end{eqnarray}
The Eq. (\ref{eqf}) can also be rewritten as the Klein-Gordon equation
for a massless scalar field in a curved (2+1)-dimensional spacetime
as follows~\cite{velten,oli,fab} 
\begin{eqnarray}
\frac{1}{\sqrt{-g}}\partial_{\mu}(\sqrt{-g}g^{\mu\nu}\partial_{\nu}\psi)=0,\label{salako14'}
\end{eqnarray}
where 
\begin{eqnarray}
f^{\mu\nu}=\sqrt{-g}g^{\mu\nu}=\frac{\rho_{0}}{c_{s}^{2}}\left[\begin{array}{ccc}
-1 & \quad-\frac{(1+\varsigma)}{2}v_{x} & \quad-\frac{(1+\varsigma)}{2}v_{y}\\
\\
-\frac{(1+\varsigma)}{2}v_{x} & \quad c_{s}^{2}-\varsigma{v}_{x}^{2} & \quad-\varsigma{v}_{x}{v}_{y}\\
\\
-\frac{(1+\varsigma)}{2}v_{y} & \quad-\varsigma{v}_{x}{v}_{y} & \quad c_{s}^{2}-\varsigma{v}_{y}^{2}
\end{array}\right].
\end{eqnarray}
So in terms of the inverse of $g^{\mu\nu}$ we obtain the effective
(acoustic) metric given in the form 
\begin{eqnarray}
g_{\mu\nu}=\sqrt{\frac{\rho_{0}}{c_{s}^{2}+v^{2}\frac{(\varsigma-1)^{2}}{4}}}\left[\begin{array}{ccc}
-(c_{s}^{2}-\varsigma v^{2}) & \quad-\frac{(1+\varsigma)}{2}v_{x} & \quad-\frac{(1+\varsigma)}{2}v_{y}\\
\\
-\frac{(1+\varsigma)}{2}v_{x} & \quad1+\frac{(\varsigma-1)^{2}}{4c_{s}^{2}}v_{y}^{2} & \quad-\frac{(\varsigma-1)^{2}}{4c_{s}^{2}}v_{x}v_{y}\\
\\
-\frac{(1+\varsigma)}{2}v_{y} & \quad-\frac{(\varsigma-1)^{2}}{4c_{s}^{2}}v_{x}v_{y} & \quad1+\frac{(\varsigma-1)^{2}}{4c_{s}^{2}}v_{x}^{2}
\end{array}\right].
\end{eqnarray}
The effective line element can be written as 
\begin{eqnarray}
ds^{2}=\sqrt{\frac{\rho_{0}}{c_{s}^{2}+v^{2}\frac{(\varsigma-1)^{2}}{4}}}\left[-(c_{s}^{2}-\varsigma v^{2})dt^{2}-(1+\varsigma)(\vec{v}\cdot d\vec{r})dt+d\vec{r}^{2}+\frac{(\varsigma-1)^{2}}{4c_{s}^{2}}\left(v_{y}dx-v_{x}dy\right)^{2}\right].
\end{eqnarray}
In polar coordinates ($\vec{v}=v_{r}\hat{r}+v_{\phi}\hat{\phi}$ and
$d\vec{r}=dr\hat{r}+rd\phi\hat{\phi}$) we have 
\begin{eqnarray}
ds^{2} & = & \tilde{\rho}\left[-\left[c_{s}^{2}-\varsigma(v_{r}^{2}+v_{\phi}^{2})\right]dt^{2}-(1+\varsigma)(v_{r}dr+v_{\phi}rd\phi)dt+(dr^{2}+r^{2}d\phi^{2})%\nonumber%
+\frac{(\varsigma-1)^{2}}{4c_{s}^{2}}(v_{\phi}dr-v_{r}rd\phi)^{2}\right],\label{salako15''}
\end{eqnarray}
where $\tilde{\rho}=\sqrt{\rho_{0}}\left[c_{s}^{2}+(v_{r}^{2}+v_{\phi}^{2})\frac{(\varsigma-1)^{2}}{4}\right]^{-1/2}$
and neo-Newtonian paramater $\varsigma=1+kn\rho_{0}^{n-1}$. At this point it is appropriate
to apply the following coordinate transformations 
\begin{eqnarray}
d\tau=dt+\frac{(1+\varsigma)v_{r}dr}{2(c_{s}^{2}-\varsigma v_{r}^{2})},\quad\quad d\varphi=d\phi+\frac{\varsigma(1+\varsigma)v_{r}v_{\phi}dr}{r(c_{s}^{2}-\varsigma v_{r}^{2})}.
\end{eqnarray}
In this way the line element can be written as 
\begin{eqnarray}\label{salako1711}
ds^{2} & = & \tilde{\rho}\Bigg\{-\Big[c_{s}^{2}-\varsigma(v_{r}^{2}+v_{\phi}^{2})\Big]d\tau^{2}+\frac{c_{s}^{2}
\left[1+(v_{r}^{2}+v_{\phi}^{2})\left(\frac{\varsigma-1}{2c_{s}}\right)^{2}\right]}{(c_{s}^{2}-\varsigma\,v_{r}^{2})}dr^{2}
-v_{\phi}(1+\varsigma)\,rd\tau d\varphi\nonumber \\
 & + & r^{2}\Big[1+v_{r}^{2}\left(\frac{\varsigma-1}{2c_{s}}\right)^{2}\Big]d\varphi^{2}\Bigg\}.
\end{eqnarray}
Now considering a static and position independent density, the {flow/fluid
velocity} is given by 
\begin{eqnarray}
\vec{v}=\frac{A}{r}\hat{r}+\frac{B}{r}\hat{\phi},\label{salako17''}
\end{eqnarray}
which is a solution obtained from the continuity equation
and the velocity potential is 
\begin{eqnarray}
\psi(r,\phi)=-A\ln r-B\phi.\label{salako18'}
\end{eqnarray}
Thus, considering $c_{s}=1$ and substituting (\ref{salako17''}) and (\ref{salako18'})
into the metric (\ref{salako1711}) we obtain the acoustic black hole in neo-Newtonian theory which is given
by 
\begin{eqnarray}
ds^{2} & = & \beta_{1}\left[-\left(1-\frac{r_{e}^{2}}{r^{2}}\right)d\tau^{2}+(1+\beta_{2})\left(1-\frac{r_{h}^{2}}{r^{2}}\right)^{-1}dr^{2}-\frac{2B\beta_{3}}{r}rd\tau d\varphi+\Big(1+\frac{\beta_{4}}{r^{2}}\Big)r^{2}d\varphi^{2}\right],\label{m-ab-nn}
\end{eqnarray}
where 
\begin{eqnarray}
\beta_{1} & = & \left(1+\beta_{2}\right)^{-1/2},\quad\quad\beta_{2}=\frac{r_{e}^{2}}{r^{2}}\left(\frac{\varsigma-1}{2}\right)^{2},\label{betas}\\
\beta_{3} & = & \frac{(1+\varsigma)}{2},\quad\quad \beta_{4}=\left(\frac{A(\varsigma-1)}{2}\right)^{2}.
\end{eqnarray}
Note that $r_{e}$ is a radius of ergo-region and $r_{h}$ is an event horizon,
i.e., 
\begin{eqnarray}
r_{e}=\sqrt{\varsigma(A^{2}+B^{2})},\quad\quad r_{h}=\sqrt{\varsigma}\vert A\vert.
\end{eqnarray}
Now, the metric (\ref{m-ab-nn}) can be written in the form of
\begin{eqnarray}
g_{\mu\nu}=\beta_{1}\left[\begin{array}{clcl}
-f & \quad\quad\quad0 & -\frac{B\beta_{3}}{r}\\
0 & \quad(1+\beta_{2}){\cal Q}^{-1} & 0\\
-\frac{B\beta_{3}}{r} & \quad\quad\quad0 & \left(1+\frac{\beta_{4}}{r^{2}}\right)
\end{array}\right],
\end{eqnarray}
and the inverse of the $g_{\mu\nu}$:
\begin{eqnarray}
g^{\mu\nu}=\frac{\beta_{1}(1+\beta_{2})}{-g}\left[\begin{array}{clcl}
-\Big(1+\frac{\beta_{4}}{r^{2}}\Big){\cal Q}^{-1} & \quad\quad0 & \quad\quad-\frac{B\beta_{3}}{r{\cal Q}}\\
0 & \quad\frac{-g{\cal Q}}{(1+\beta_{2})^{2}} & \quad\quad0\\
-\frac{B\beta_{3}}{r{\cal Q}} & \quad\quad0 & \quad\quad\frac{f_{1}}{{\cal Q}}
\end{array}\right],\label{metrinv}
\end{eqnarray}
where } 
\begin{eqnarray}
f_{1} & = & 1-\frac{r_{e}^{2}}{r^{2}},\quad\quad{\cal Q}=1-\frac{r_{h}^{2}}{r^{2}},\\
-g & = & \frac{(1+\beta_{2})}{{\cal Q}}\left[\left(1+\frac{\beta_{4}}{r^{2}}\right)f_{1}+\frac{B^{2}\beta_{3}^{2}}{r^{2}}\right].
\end{eqnarray}

Next we consider a general symmetric
metric form (\ref{m-ab-nn}) for $B=0$,

\begin{eqnarray}
ds^{2} & = & \beta_{1}\left[-f(r)\,d\varsigma^{2}+\frac{(1+\beta_{2})}{f(r)}dr^{2}+\Big(1+\frac{\beta_{4}}{r^{2}}\Big)r^{2}d\varphi^{2}\right],\label{spherically}
\end{eqnarray}
with 
\begin{eqnarray}
f(r) & = & \left(1-\frac{r_{h}^{2}}{r^{2}}\right)\\
 & = & \left(1-\frac{\varsigma\,A^{2}}{r^{2}}\right),
\end{eqnarray}
or we can write it in more compact form as follows:

\be
ds^{2}=-\mathcal{F}\,d\varsigma^{2}+\mathcal{G}dr^{2}+\mathcal{H} d\varphi^{2}, \label{4}
\ee
where $\mathcal{F}=\sqrt{\beta_{1}f(r)}$ , $\mathcal{G}=\sqrt{\frac{\beta_{1}(1+\beta_{2})}{f(r)}}$
and $\mathcal{H}=\sqrt{\beta_{1} r^2 \Big(1+\frac{\beta_{4}}{r^{2}}\Big)}$.

\section{Construction of Thin-Shell Wormholes in Neo-Newtonian theory }\label{section4}
In this section, we construct the thin-shell wormholes in neo-Newtonian theory by using the metric \eqref{4}. To construct the wormhole, we use the cut and paste technique \cite{baji,bilic,fagnic,visser}. First, we choose two identical regions 
\begin{eqnarray}
M^{(\pm)}=\left\lbrace r^{(\pm)}\geq a,\,\,a>r_{h}\right\rbrace ,
\end{eqnarray}
in which $a$ is chosen to be greater than the event horizon $r_{h}$.
If we now paste these regular regions at the boundary hypersurface
$\Sigma^{(\pm)}=\left\lbrace r^{(\pm)}=a,a>r_{H}\right\rbrace $,
then we end up with a complete manifold $M=M^{+}\bigcup M^{-}$. In
accordance with the Darmois\textendash Israel formalism the coordinates
on $M$ can be choosen as $x^{\alpha}=(t,r,\theta,\phi)$. On the
other hand for the coordinates on the induced metric $\Sigma$ we
write $\xi^{i}=(\tau,\theta,\phi)$.
Finally for the parametric equation on the induced metric $\Sigma$
we write 
\begin{eqnarray}
\Sigma:R(r,\tau)=r-a(\tau)=0.
\end{eqnarray}

Note that in order to study the dynamics of the induced metric $\Sigma$,
in the last equation we let the throat radius of the wormhole to be
time dependent by incorporating the proper time on the shell i.e.,
$a=a(\tau)$. For
the induced metric we have the spacetime on the shell\begin{eqnarray}
\mathrm{d}s_{\Sigma}^{2}=-\mathrm{d}\tau^{2}+a(\tau)^{2}\mathrm{d}\phi^{2}.\label{metric}
\end{eqnarray}

The junction conditions on $\Sigma$ reads 
\begin{eqnarray}
{S^{i}}_{j}=-\frac{1}{8\pi}\left(\left[{K^{i}}_{j}\right]-{\delta^{i}}_{j}\,K\right).
\end{eqnarray}

Note that in the last equation ${S^{i}}_{j}=diag(-\sigma,p)$ is the
energy momentum tensor on the thin-shell, on the other hand $K$,
and $[K_{ij}]$, are defined as $K=trace\,[{K^{i}}_{i}]$ and $[K_{ij}]={K_{ij}}^{+}-{K_{ij}}^{-}$,
respectively. Keeping this in mind, we can go on by writing the expression
for the extrinsic curvature ${K^{i}}_{j}$ as follows 
\begin{eqnarray}
K_{ij}^{(\pm)}=-n_{\mu}^{(\pm)}\left(\frac{\partial^{2}x^{\mu}}{\partial\xi^{i}\partial\xi^{j}}+\Gamma_{\alpha\beta}^{\mu}\frac{\partial x^{\alpha}}{\partial\xi^{i}}\frac{\partial x^{\beta}}{\partial\xi^{j}}\right)_{\Sigma}.
\end{eqnarray}

The unit vectors ${n_{\mu}}^{(\pm)}$, which are normal to $M^{(\pm)}$
are choosen as 
\begin{eqnarray}
n_{\mu}^{(\pm)}=\pm\left(\left\vert g^{\alpha\beta}\frac{\partial R}{\partial x^{\alpha}}\frac{\partial R}{\partial x^{\beta}}\right\vert ^{-1/2}\frac{\partial R}{\partial x^{\mu}}\right)_{\Sigma}.
\end{eqnarray}

\be
n_t=\mp\dot a\sqrt{\mathcal{G} \mathcal{F}},
\ee
\be 
n_r=\pm\sqrt{\mathcal{G}[1+{\dot a}^2\mathcal{G}]}.
\ee
Then, the extrinsic curvature is given by \cite{baji,bilic,fagnic,visser}
\be
K^\pm_{\tau \tau}=\mp\frac{\sqrt{\mathcal{G}}}{2\sqrt{1+{\dot a}^2\mathcal{G}}}\left\{ 2\ddot a+{\dot a}^2\left[\frac{\mathcal{F}'}{\mathcal{F}}+ \frac{\mathcal{G}'}{\mathcal{G}}\right]+\frac{\mathcal{F}'}{\mathcal{F}\mathcal{G}} \right\},
\ee
\be
K^\pm_{\theta\theta}=\pm\frac{\mathcal{H}'}{2\mathcal{H}}\sqrt{\frac{1+{\dot a}^2\mathcal{G}}{\mathcal{G}}},
\ee

Using the definitions $[K_{_{{i}{j}}}]\equiv K_{_{{i}{j}}}^{+}-K_{_{{i}{j}}}^{-}$,
and $K=tr[K_{{i}{j}}]=[K_{\;{i}}^{{i}}]$, and the surface stress\textendash energy
tensor $S_{_{{i}{j}}}={\rm diag}(\sigma,p)$ it follows the Lanczos
equations on the shell 
\begin{eqnarray}
-[K_{{i}{j}}]+Kg_{{i}{j}}=8\pi S_{{i}{j}}.\label{eq7}
\end{eqnarray}

Note that for a given radius $a$, the energy density on the shell
is $\sigma$, while the pressure $p=p_{{\theta}}$.
If we now combine the above results for the surface density \cite{baji,bilic,fagnic,visser}
\begin{eqnarray}
\sigma=-\frac{1}{8\pi}\frac{\mathcal{H}'}{\mathcal{H}}\sqrt{\frac{1+{\dot a}^2\mathcal{G}}{\mathcal{G}}}
,\label{25}
\end{eqnarray}
and the surface pressure 
\begin{eqnarray}
p=\frac{1}{8\pi}\sqrt{\frac{\mathcal{G}}{1+{\dot a}^2\mathcal{G}}}\left\{ 2\ddot a +{\dot a}^2\left[\frac{\mathcal{F}'}{\mathcal{F}}+ \frac{\mathcal{G}'}{\mathcal{G}}\right]+\frac{\mathcal{F}'}{\mathcal{F}\mathcal{G}}\right\}.\label{26}
\end{eqnarray}

Since we are going to study the wormhole stability at a static configuration
we need to set $\dot{a}=0$, and $\ddot{a}=0$. For the surface density
in static configuration it follows that 

\begin{eqnarray}
\sigma_{0}=-\frac{1}{8\pi}\frac{\mathcal{H}'}{\mathcal{H}\sqrt{\mathcal{G}}},\label{27}
\end{eqnarray}
and similarly the surface pressure 
\begin{eqnarray}
p_{0}=\frac{1}{8\pi}\frac{\mathcal{F}'}{\mathcal{F}\sqrt{\mathcal{G}}}.\label{28}
\end{eqnarray}

It's obvious from Eq. \eqref{27} that the surface density is negative,
i.e. $\sigma_{0}<0$, which implies that the weak and dominant energy
conditions are violated. 

\section{Stability Analysis}\label{section5}

In this section we are going to analyze the stability of the WH. Starting
from the energy conservation it follows that \cite{baji,bilic,fagnic,visser}

\be
-\nabla_i S^i_j = \left[T_{\alpha \beta} \frac{\partial x^\alpha}{\partial \xi^j}
  n^\beta \right],
\label{ceq}
\ee

\begin{eqnarray}
\frac{d}{d\tau}\left(\sigma\mathcal{A}\right)+p\frac{d\mathcal{A}}{d\tau}=-\frac{\dot a \sigma {\cal A}}{2}\left[\frac{\mathcal{F}'}{\mathcal{F}}+ \frac{\mathcal{G}'}{\mathcal{G}}+\frac{\mathcal{H}'}{\mathcal{H}}-\frac{2\mathcal{H}''}{\mathcal{H}'}\right],\label{34}
\end{eqnarray}
where $\mathcal{A}$ is the area of the wormhole throat. By replacing
$\sigma(a)$ we can find the equation of motion as follows 
\begin{eqnarray}
\dot{a}^{2}=-V(a),\label{35}
\end{eqnarray}
with the potential 
\begin{eqnarray}
V(a)=\frac{1}{\mathcal{G}}-\left[8\pi \lambda \frac{\mathcal{H}}{\mathcal{H}'}\right]^2.\label{36}
\end{eqnarray}

In order to investigate the stability of WH let us expand the potential
$V(a)$ around the static solution by writing 
\begin{eqnarray}
V(a)=V(a_{0})+V^{\prime}(a_{0})(a-a_{0})+\frac{V^{\prime\prime}(a_{0})}{2}(a-a_{0})^{2}+O(a-a_{0})^{3}.\label{37}
\end{eqnarray}

The second derivative of the potential is

\begin{eqnarray}
V''(a_0)&=& -\frac{\mathcal{F} \mathcal{F}' \mathcal{G}'+2 \mathcal{G} \left\{ \mathcal{F}'^2-\mathcal{F} \mathcal{F}''\right\} }{2 \mathcal{F}^2 \mathcal{G}^2} + \psi^{\prime} \frac{\mathcal{H} \left[ 2 \mathcal{G} \mathcal{H}''-\mathcal{G}' \mathcal{H}'\right] -2 \mathcal{G} \mathcal{H}'^2}{2 \mathcal{G}^2 \mathcal{H}^2}. \label{v2s} 
\end{eqnarray}
where we have introduced $\psi^{\prime}=p'/\sigma'$.

The wormhole is stable if and only if $V^{\prime\prime}(a_{0})>0$.
The equation of motion of the throat, for a small perturbation becomes
\begin{eqnarray}
\dot{a}^{2}+\frac{V^{\prime\prime}(a_{0})}{2}(a-a_{0})^{2}=0.
\end{eqnarray}

Noted that for the condition of $V^{\prime\prime}(a_{0})\geq0$, WH
is stable where the motion of the throat is oscillatory with angular
frequency $\omega=\sqrt{\frac{V^{\prime\prime}(a_{0})}{2}}$. In this
work we are going to use five different models for the fluid to explore
the stability analysis; the LBF \cite{kuf2,varela}, the CF \cite{cg1,cg2,GCCG,GCCG1}, the LogF \cite{ali1,ali2} and finally PF  \cite{sarkar}.

\subsection{Stability analysis of ATW via the LBF}

In our first case, we choose the LBF with the equation
of state given by \cite{kuf2,varela}
\begin{eqnarray}
\psi=\omega\sigma,
\end{eqnarray}
it follows that 
\begin{eqnarray}
\psi^{\prime}(\sigma_{0})=\omega.
\end{eqnarray}

\begin{figure}[h!]
\includegraphics[width=0.40\textwidth]{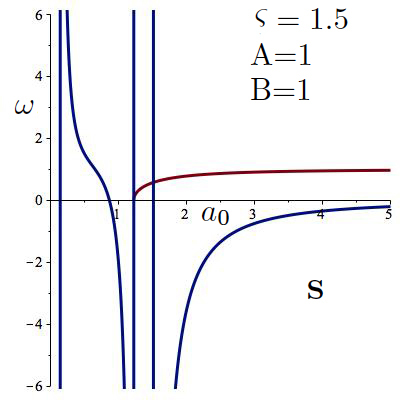} %
\includegraphics[width=0.50\textwidth]{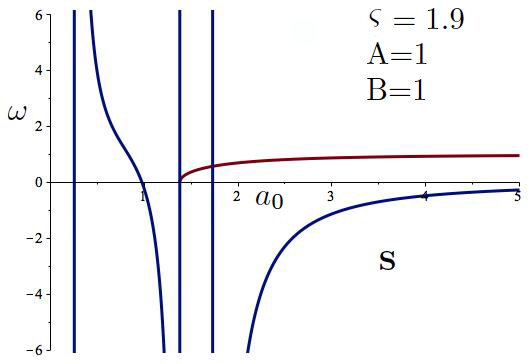}
\includegraphics[width=0.40\textwidth]{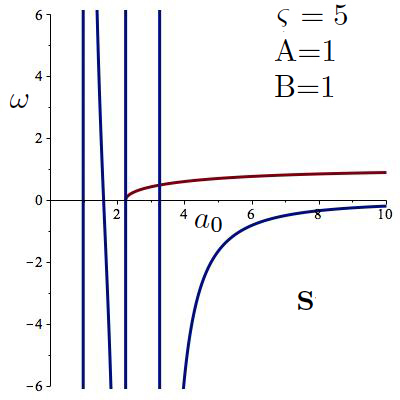}
\includegraphics[width=0.40\textwidth]{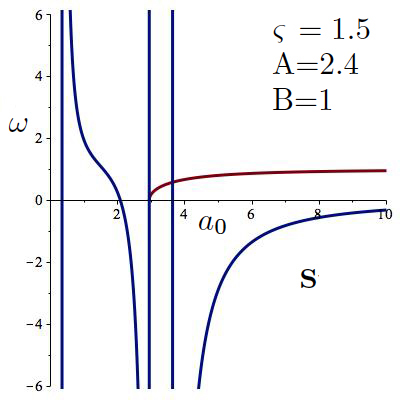}
\caption{ \textit{Here we plot the stability regions  for the LBF as a function of $\omega$ and radius of the throat $a_{0}$.  } }
\label{fig1}
\end{figure}

Note that $\omega$ is a constant parameter. In order to see more
clearly the stability we show graphically the dependence of $\omega$
in terms of $a_{0}$ for different values of the parameter $\varsigma$, $A$ and $B$
in Fig.\ref{fig1}.

\subsection{Stability analysis of ATW via the CF}

According to the CF, we can model the fluid with the following equation of state \cite{cg1,cg2,GCCG,GCCG1}
\begin{equation}
\psi=\omega \left(\frac{1}{\sigma}-\frac{1}{\sigma_0}\right)+p_0,
\end{equation}
to find
\begin{equation}
\psi^{\prime}(\sigma_{0})=-\frac{\omega}{\sigma_0^{2}}.
\end{equation}

\begin{figure}[h!]
\includegraphics[width=0.45\textwidth]{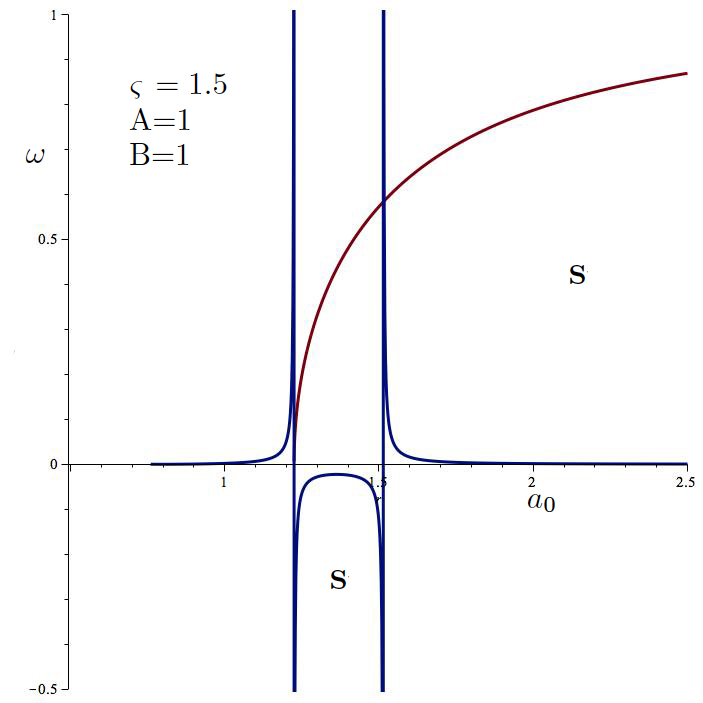} %
\includegraphics[width=0.45\textwidth]{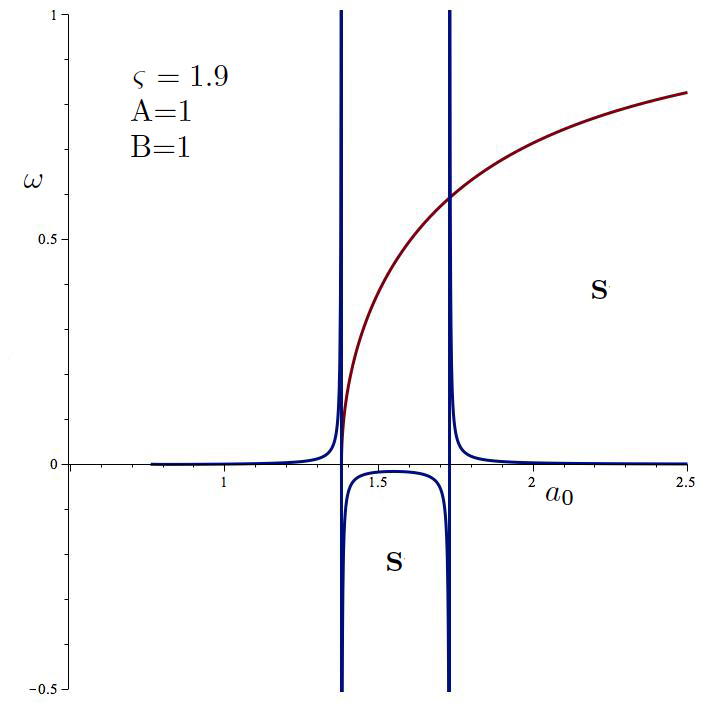}
\includegraphics[width=0.40\textwidth]{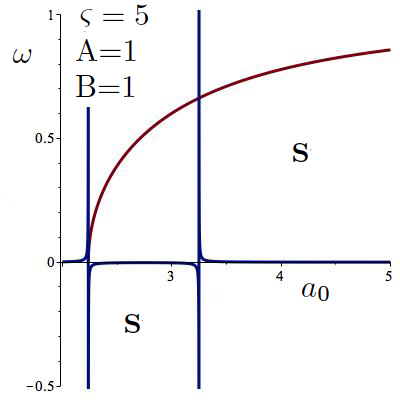}
\includegraphics[width=0.40\textwidth]{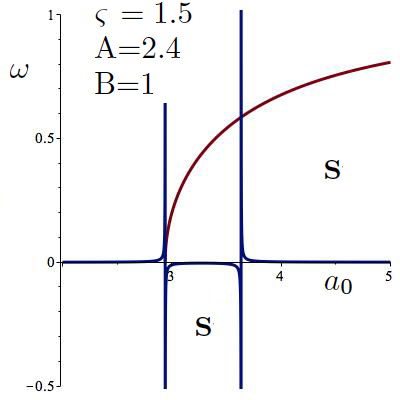} %
\caption{\textit{Here we plot the stability regions via the CF as a function of $\omega$ and radius of the throat
$a_{0}$.  } }
\label{fig2}
\end{figure}

To see the stability regions let us show graphically the dependence of $\omega$ in terms of $a_{0}$ for
different values of the parameter $\varsigma$, $A$ and $B$, given in Fig.\ref{fig2}. 

\subsection{Stability analysis of ATW via the LogF}

Our next example is the LogF \cite{ali1,ali2}, with the equation of state 
\begin{equation}
\psi=\omega \ln \left(\frac{\sigma}{\sigma_0}\right)+p_0,
\end{equation}
then
\begin{equation}
\psi^{\prime}(\sigma_{0})=\frac{\omega}{\sigma_0}.
\end{equation}

\begin{figure}[h!]
\includegraphics[width=0.45\textwidth]{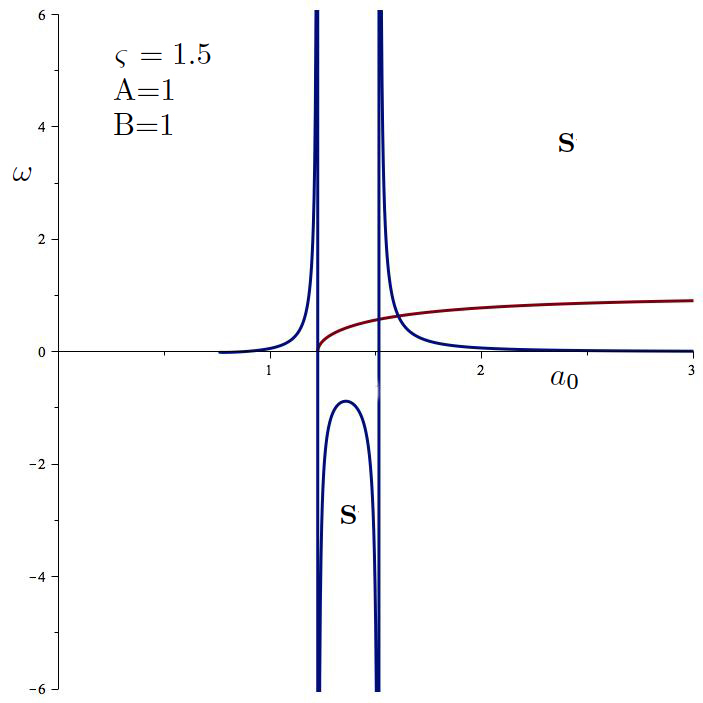} %
\includegraphics[width=0.45\textwidth]{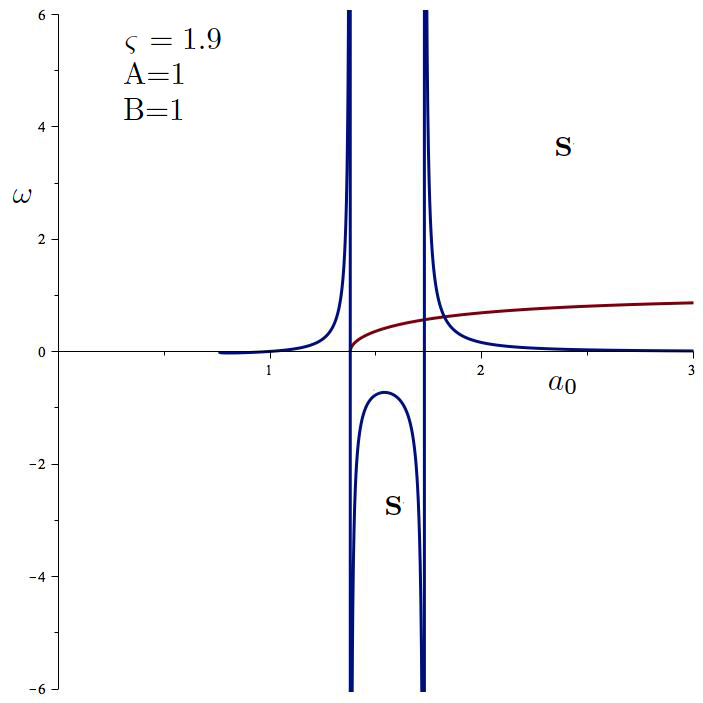} %
\includegraphics[width=0.40\textwidth]{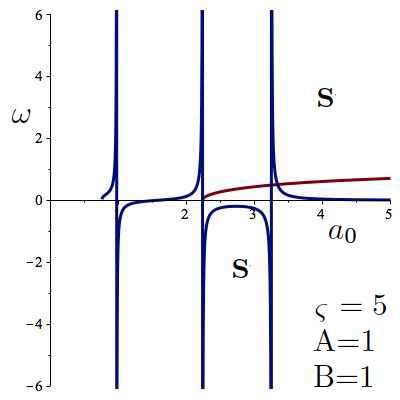} %
\includegraphics[width=0.40\textwidth]{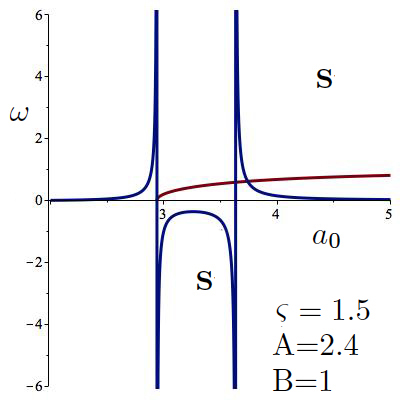}
\caption{ \textit{The stability regions as a function via the LogF of $\omega$ and radius of the throat $a_{0}$, in which we have choosen three different values $\varsigma$, $A$ and $B$.} }
\label{fig3}
\end{figure}

For detailed information we can show graphically the dependence of $\omega$ in terms of $a_{0}$
by choosing different values of the parameter $\varsigma$, $A$ and $B$, in Fig.\ref{fig3}. 

\subsection{Stability analysis of ATW via PF}

The equation of state for the fluid according to the PF can be written as \cite{sarkar,poli}
\begin{eqnarray}
\psi=\omega\sigma^{\gamma},
\end{eqnarray}

It follows that 
\begin{eqnarray}
\psi^{\prime}(\sigma_{0})=\omega\,\gamma\,\sigma_{0}^{\gamma-1}.
\end{eqnarray}

\begin{figure}[h!]
\includegraphics[width=0.45\textwidth]{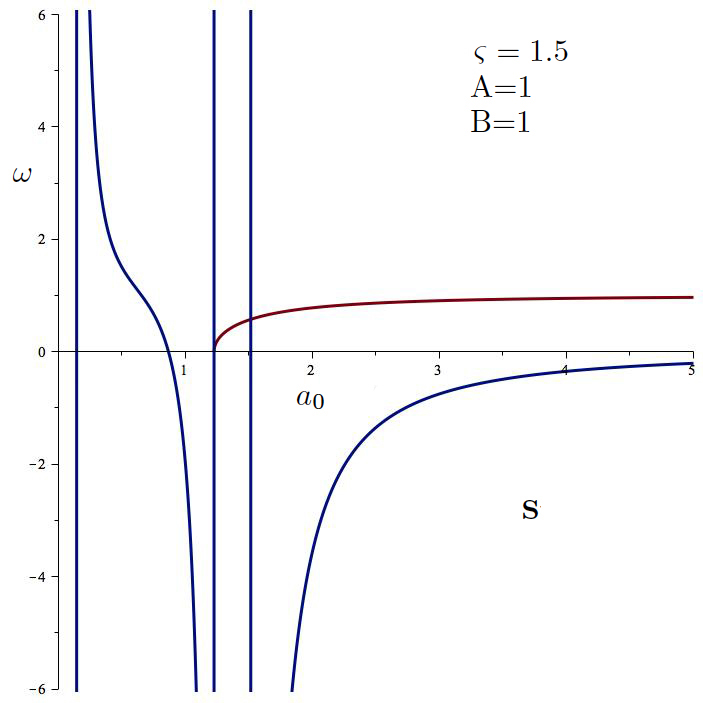} %
\includegraphics[width=0.45\textwidth]{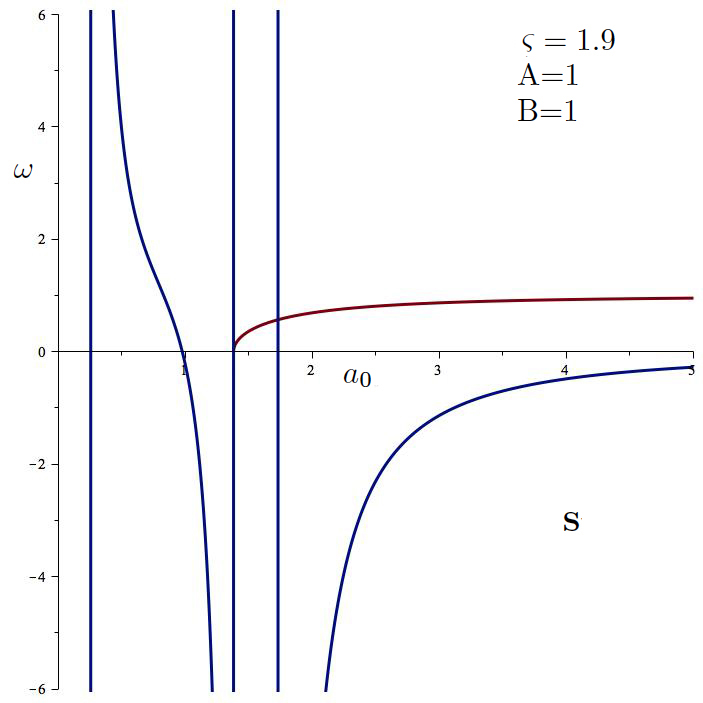}
\includegraphics[width=0.40\textwidth]{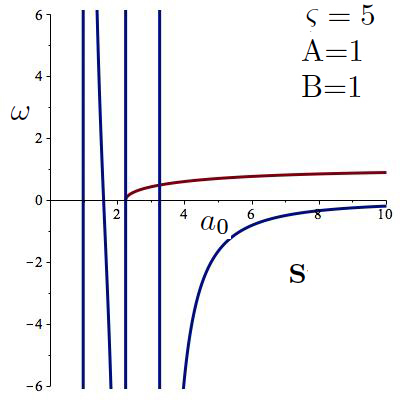}
\includegraphics[width=0.40\textwidth]{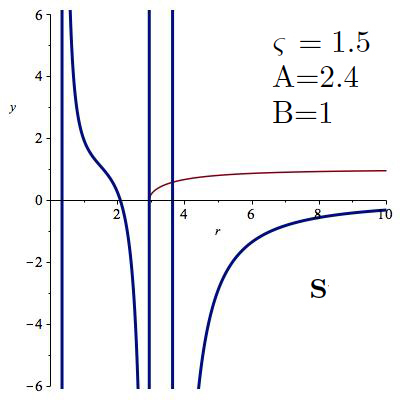}
\caption{ \textit{Here we plot the stability regions via the PF as a function of $\omega$ and 
radius of the throat $a_{0}$ for the parameter $\gamma=1.1$. } }
\label{fig4}
\end{figure}

For detailed information we plot $\omega$ in terms of $a_{0}$ by
choosing different values of the parameter $\varsigma$, $A$ and $B$, as shown in Fig.\ref{fig4}.

\section{Conclusion}\label{section6}

In this paper, we constructed a new ATW
in the context of neo-Newtonian hydrodynamics, which is a modification of the usual Newtonian theory that correctly
incorporates the effects of pressure. We used a cut-and-paste technique to join together two
regular regions, and then we computed the analog surface density and surface pressure of the fluid.  
The stability analyses were carried out using an LBF, CF, LogF, and
PF to show that the ATW can be stable if
one chooses suitable values of the parameters $\varsigma$, $A$, and $B$. In Fig. 1, after we chose specific values of $\varsigma$ as $1.5$, $1.9$, and $5$ with $A=B=1$ and $\varsigma=1.5$ with $A=2.4$ and $B=1$, we showed that the stability region (S) for the ATW is supported with the LBF. The sizes of the stability regions decrease with increasing values of $\varsigma$ and $A$. In Fig. 2, we chose values of $\varsigma$ as $1.5$, $1.9$, and $5$ with $A=B=1$ and $\varsigma=1.5$ with $A=2.4$ and $B=1$ to show the effect of the CF on the ATW. Then, in Fig. 3, we chose the same parameters again, but we used different fluids, such as the LogF, to show that the stability regions change with varying values of the parameters $\varsigma$ and $A$. Lastly, we used the PF to show the stability regions, using the same parameters. Increasing values of the neo-Newtonian parameter $\varsigma$ decrease the sizes of the stability regions. Thus, we noted that the effects of pressure influence the stability of the model. We showed that fluids with negative energy are required at the
throat to keep the wormhole stable, leading us to conclude that $\varsigma$ is the most critical factor for the existence of a stable ATW. 

\begin{acknowledgments}
This work was supported by the Chilean FONDECYT Grant No. 3170035 (A\"{O}). We would like to thank the anonymous reviewers for their useful comments and suggestions which helped us to improve the paper.
\end{acknowledgments}

\end{document}